\documentclass[article]{elsarticle}
\usepackage[T1]{fontenc}
\usepackage{caption}
\usepackage{subcaption}
\usepackage{lineno,hyperref}
\usepackage{amsmath}
\usepackage{amssymb}
\usepackage{mathtools}
\usepackage{tablefootnote}
\usepackage{soul}
\usepackage{ulem}
\usepackage{color}
\usepackage{float}
\usepackage{makecell}
\usepackage[usenames,dvipsnames,svgnames,table]{xcolor}
\usepackage{indentfirst}
\usepackage{leftidx}
\usepackage{slashed}

\begin{document}
	\begin{frontmatter}
		\title{Distinguishing Dark Energy Models with Neutrino Oscillations}
		
		\author[ICC,UB]{Ali Rida Khalifeh\corref{cor1}}
		\ead{ark93@icc.ub.edu}
		\cortext[cor1]{Corresponding author}
		
		\author[ICC,ICREA]{Raul Jimenez}
		\ead{raul.jimenez@icc.ub.edu}

		\address[ICC]{ICC, University of Barcelona, Marti  i Franques, 1, E-08028 Barcelona, Spain.}
		\address[UB]{Dept. de  Fisica Cuantica y Astrofisica, University of Barcelona, Marti  i Franques 1, E-08028 Barcelona, Spain.}
		\address[ICREA]{ICREA, Pg. Lluis Companys 23, Barcelona, E-08010, Spain.}

		\begin{abstract}
			Dark Energy models are numerous and distinguishing between them is becoming difficult. However, using distinct observational probes can ease this quest and gives better assessment to the nature of Dark energy. To this end, the plausibility of neutrino oscillations to be a probe of Dark Energy models is investigated. First, a generalized formalism of neutrino (spinor field) interaction with a classical scalar field in curved space-time is presented. This formalism is then applied to two classes of Dark Energy models in a flat Friedman-Lema\^itre-Robertson-Walker metric: a Cosmological Constant and scalar field Dark Energy coupled to neutrinos. By looking at the neutrino oscillation probability's evolution with redshift, these models can be distinguished, for certain neutrino and scalar field coupling properties. This evolution could be traced by neutrino flux measurements in future underground, terrestrial or extraterrestrial, neutrino telescopes which would assess probing Dark Energy models with this technique.
		\end{abstract}
	\end{frontmatter}

	
	\section{Introduction}
	\label{Intro.}
	
	Since the discovery of the accelerated expansion of the Universe~\cite{Perlmutter:acelerating2,Riess:acelerating1}, one of the most interesting open questions in Astrophysics and Cosmology is to understand if Dark Energy(DE) is dynamic, or instead strictly a constant.  Indeed, if DE was shown to be dynamical, this would be a major revolution, as it would indicate a great deal of new physics. However, recent observational constraints indicate that DE is consistent with a cosmological constant, with a few percent uncertainty~\cite{Planck2018}. Current and upcoming cosmological surveys, such as DESI~\cite{DESI} and Euclid~\cite{EUCLID2} will decrease this level of uncertainty to the \% level~\cite{EUCLID1}. Nevertheless, theoretical arguments have been presented over the years in favor of a dynamical DE, due to fundamental issues accompanied by a constant one, such as the coincidence problem~\cite{Velten:Coincidence}, see also~\cite{Wang:Coincidence,Weinberg:LambdaProblem,Weinberg:LambdaProblem2}.
	
	In order to lift this dilemma, one could combine several probes and techniques to constraint DE models. In addition to the already mentioned probes, as well as Gravitational Waves surveys~\cite{LIGO-VIRGO,LIGO-DE,Creminelli-DE}, looking at neutrinos could open a new window to the nature of DE. Several Cosmological probes have been used to constrain neutrino properties in the context of a flat Friedman-Lema\^itre-Robertson-Walker(FLRW) Universe~\cite{Lesgourgues:neutrinobiblia,Betti:Ptolemy,Takada:freestreaming,Jimenez:freestreaming2,Moresco:wowaconstrain,2007PhLB..655..201W}. However, neutrinos have been mostly considered as classical point particles, rather than quantum spinor fields traveling in curved spacetime, which could provide novel insights for both neutrinos and DE. An interaction between spinor and scalar fields is sensible to our spacetime's curved geometry, which could leave observational imprints. This shows the advantage of using Cosmology to understand properties of the Universe, for it can thus give information on both Gravity and neutrinos.
	
	Studying neutrinos as quantum spinors in curved spacetime have been done in several theoretical contexts, such as near Schwarzschild Blackholes~\cite{Cardall-Fuller}, in extended theories of gravity~\cite{Buoninfante:2019der} or to derive fundamental uncertainty relations~\cite{Time-EnergyUncertainty,Blasone:2018ktu}. More specifically to dynamical DE, neutrinos have been investigated in the context of mass-varying neutrinos~\cite{PhysRevD.73.083515,Fardon_2004,Kaplan-Nu,Mohseni,MohseniSadjadi:2017jne}, pseudo-Dirac particles~\cite{Pseudo-Dirac1,Pseudo-Dirac2} or Lorentz/$CPT$ violating theories~\cite{Kamionkowski_Nu_Lorentz,Nu_DE_Data}, where $CPT$ stands for Charge Conjugation($C$), Parity($P$) and Time reversal($T$) symmetries. It would be interesting therefore to expand on these works to produce an observational trace of this kind of interactions.
	
	As a first step along this way, one of us proposed a model, called DE${}_{\nu}$, in which a scalar field is ``frozen'' in place via an interaction with neutrino~\cite{2018PDU....20...72S}. This model, by construction, mimics a cosmological constant from the point of view of cosmological observables (expansion history and perturbations) and thus it does not leave any significant imprint in these classical observables. We then expanded upon that work~\cite{Khalifeh:2020bdg}, and looked at DE${}_{\nu}$ in the context of quantum spinors in curved spacetime, in addition to beyond Standard Model scalar-spinor interactions.
	
	In this work, we further develop ref.~\cite{Khalifeh:2020bdg}, as well as the works previously mentioned, to produce an observable that could be measured experimentally, which can then differentiate between DE models. We look at a more general massive spinor-scalar field interaction(section~\ref{Gen.Form.}), and then derive a generalized formula for the oscillation probability in an arbitrary spacetime(section~\ref{Nu_Oscill}). Although a scalar field DE scenario does not include all types of DE, nevertheless it incorporates a large class of DE models, including scalar-tensor theories of gravity such as Horndeski~\cite{Horndeski:1974wa,Kobayashi:2019hrl}. That is the reason  why we focus on such an interaction here. Afterwards, we specify to cosmological constant DE(section~\ref{Sec:LCDM}), what is known as $\Lambda$CDM model, and quintessence~\cite{Peebled_Bharat,Tsujikawa:2013fta} with a neutrino-scalar interaction as presented in the DE${}_{\nu}$ model(section~\ref{Sec:Quintessence}). For the latter, we look at various neutrino and DE${}_{\nu}$ properties and compare them to the former model. We finish by presenting a summary and future prospects(section~\ref{Sec:Conc.}).

	
	Throughout the paper, we use units in which $\hbar=c=1$, where $c$ is the speed of light and $\hbar$ is the reduced Planck constant. Moreover, the metric signature we use is the mostly positive one, $(-,+,+,+)$, and Greek indices will be used for spacetime coordinates $(0,1,2,3)$, while Latin ones are dedicated for spatial coordinates only, $(1,2,3)$. In addition, for neutrino states notations, we use Greek and Latin indices to describe flavor and mass states, respectively.
	
	
	\section{Generalized Formalism}
	\label{Gen.Form.}
	
	Let us consider a general interaction of a spinor field, $\psi$, with a classical scalar field, $\varphi$, in curved spacetime with metric $g_{\mu\nu}$. This is described by the following action:
	\begin{align}
	S=\int d^4x\sqrt{-g}\bigg[&\frac{1}{2}R-\frac{1}{2}\mathcal{D}_{\mu}\mathcal\varphi{D}^{\mu}\varphi-V(\varphi)\nonumber\\
	&+i\big(\bar{\psi}\gamma^{\mu}\mathcal{D}_{\mu}\psi-\mathcal{D}_{\mu}\bar{\psi}\gamma^{\mu}\psi\big)-2m\bar{\psi}\psi+\zeta\Theta\bigg].
	\label{Gen.Action}
	\end{align}
	In here, $g$ is the determinant of $g_{\mu\nu}$ and $R$ is the Ricci scalar, the trace of the Ricci tensor $R_{\mu\nu}$. Moreover, $\mathcal{D}_{\mu}$ is a generalized covariant derivative for fields with different spins in curved spacetime (see~Ref.\cite{Birrell:1982ix,Lanzagorta,Mukhanov:2007zz} for more details on quantum fields in curved backgrounds). For example, when acting a particle of spin 0, $\mathcal{D}_{\mu}$ reduces to $\partial_{\mu}$, the usual partial derivative in flat spacetime. We will see shortly the form it takes when acting on spinors. Another term that appears in~\eqref{Gen.Action} is the scalar field potential $V(\varphi)$, which describes self interactions of $\varphi$. Furthermore, $\gamma^{\mu}=\{\gamma^0, \gamma^1, \gamma^2, \gamma^3\}$ are the four Dirac matrices, $\bar{\psi}=\psi^{\dagger}\gamma^0$, with $\psi^{\dagger}$ being the complex transpose of $\psi$, and $m$ is the spinor field's mass. Finally, $\zeta$ is the coupling constant for the general interaction term between $\psi$ and $\varphi$, $\Theta\big(\psi, \bar{\psi}, \varphi, X^{\mu}_{\psi}, X^{\mu}_{\bar{\psi}}, X^{\mu}_{\varphi}\big)$, with $X^{\mu}_{\psi}\big(X_{\bar{\psi}}^{\mu}\big)=\mathcal{D}^{\mu}\psi\big(\mathcal{D}^{\mu}\bar{\psi}\big)$ and $X_{\varphi}^{\mu}=\partial^{\mu}\varphi$. 
	
	By setting the variational derivative of the action~\eqref{Gen.Action} with respect to(w.r.t) $\bar{\psi}$ to 0, we get the modified Dirac equation for $\psi$, i.e.
	\begin{equation}
	\frac{1}{\sqrt{-g}}\frac{\delta S}{\delta\bar{\psi}}=0\quad\Rightarrow\quad \big(i\gamma^{\mu}\mathcal{D}_{\mu}-m\big)\psi=-\frac{\zeta}{2}\bigg(\frac{\partial\Theta}{\partial\bar{\psi}}-\mathcal{D}^{\mu}\frac{\partial\Theta}{\partial X_{\bar{\psi}}^{\mu}}\bigg)\equiv -\frac{\zeta}{2}\frac{\delta\Theta}{\delta\bar{\psi}}.
	\label{E.of.M}
	\end{equation}
	At this stage, one could say that the interaction is of the most general form, for neither $\Theta$ nor the metric have been specified. However, as we are considering a more phenomenological approach to the question at hand, it would be more useful to look for a practical form of the variational derivative of $\Theta$ w.r.t $\bar{\psi}$. This would allow us to calculate observables that could be eventually measured by experiments. Moreover, it would prove useful to divide the interaction term into flavor-invariant and flavor-dependent parts, to study how each would affect the transition probability from one flavor state to another. Intuitively, one would expect that the former should not modify this flavor oscillation probability, since by definition the latter is a transition between flavors. However, as we will see in the next section, this is not always the case.
	
	There are several ways in which one can implement these considerations. For instance, in order to have the right hand side(r.h.s) of~\eqref{E.of.M} mathematically and dimensionally consistent with its left hand side(l.h.s), one possibility is: 
	\begin{equation}
	-\frac{\zeta}{2}\frac{\delta\Theta}{\delta\bar{\psi}}=\big(\xi \gamma^{\mu}F_{\mu}(\varphi,X_{\varphi}^{\mu})+\xi_f G(\varphi, X_{\varphi}^{\mu})\big)\psi.
	\end{equation}
	The first term on the r.h.s is a global interaction in flavor space, i.e. it couples to all flavors with the same strength $\xi$. On the other hand, the second term is a flavor-specific term, with the coupling strength $\xi_f$ depending on which flavor is being considered.	Another term that could be added is a kinetic coupling, such as $\xi F(\varphi,X_{\varphi}^{\mu})\gamma^{\mu}\mathcal{D}_{\mu}\psi$. However, as such a term could produce effects on other Cosmological observables(see appendix in~\cite{Khalifeh:2020bdg}), we will not be considering it here. Finally, for the purpose we seek of studying $\Lambda$CDM and quintessence(in the context of DE${}_{\nu}$), it turns out that having 
	\begin{equation}
	-\frac{\zeta}{2}\frac{\delta\Theta}{\delta\bar{\psi}}=\big(\xi F(\varphi,X_{\varphi}^{\mu})+\xi_f\gamma^{\mu}G_{\mu}(\varphi, X_{\varphi}^{\mu})\big)\psi.
	\label{Gen.Inter.}
	\end{equation}
	is more useful, and will therefore be used in the following sections
	
	\section{Neutrino Oscillation}
	\label{Nu_Oscill}
	In this section, for simplicity, we will be studying two-flavor neutrino oscillation in curved space-time, although one could generalize the analysis to three-flavor oscillations (see Ref.~\cite{Cardall-Fuller,Buoninfante:2019der,Time-EnergyUncertainty} for more details on neutrino oscillations in curved space-time). In addition, a more stringent study of neutrino oscillations in curved backgrounds would rely on the full quantum field theoretic treatment (see Ref.~\cite{Blasone:2018ktu,Capolupo} and references within for further details). However,  for the purpose of studying neutrino interaction with DE, the quantum mechanical treatment presented here is sufficient for comparison with observations. We will look at the quantum field theoretic treatment for both fields in future works.
	\subsection{Transition Amplitude's Evolution}
	The first step in studying neutrino oscillations is to expand a state of flavor $\alpha$, $|\nu_{\alpha}\rangle$, in terms of mass eigenstates, $|\nu_j\rangle$:
	\begin{equation}
	|\nu_{\alpha}(\lambda)\rangle=\sum_{j=1,2}^{ }U_{\alpha j}e^{i\Phi(\lambda)}|\nu_j\rangle,
	\label{State_Expansion}
	\end{equation}  
	where $\lambda$ is the monotonically increasing affine parameter along the neutrino world-line and $U_{\alpha j}$ is the two-flavor mixing matrix, given by:
	\begin{align}
	U &= \begin{pmatrix}
	\cos\theta & \sin\theta \\
	-\sin\theta & \cos\theta
	\end{pmatrix}
	\label{Mixing_Matrix}
	\end{align}
	with $\theta$ being the mixing angle. Moreover, $\Phi(\lambda)$ is the mass eigenstate's evolution operator~\cite{Cardall-Fuller}:
	\begin{equation}
	\Phi(\lambda)=\int_{\lambda_0}^{\lambda}P_{\mu}\frac{dx^{\mu}}{d\lambda'}d\lambda'
	\label{Evolution_Operator}
	\end{equation}
	where $P^{\mu}$ is the 4-momentum operator, $dx^{\mu}/d\lambda$ is a null vector tangent to the neutrino world-line, and $\lambda_0$($\lambda$) is the affine parameter's value at the observer(source).
	
	One can see that eq.~\eqref{State_Expansion} is a solution for the Schr\"odinger like equation
	\begin{equation}
	i\frac{d}{d\lambda}|\nu_{\alpha}(\lambda)\rangle=\Phi(\lambda)|\nu_{\alpha}(\lambda)\rangle,
	\label{Diff_Equ_Evolution}
	\end{equation}
	and therefore the transition amplitude between states $\alpha$ and $\beta$,
	\begin{equation}
	\Psi_{\alpha\beta}=\langle\nu_{\beta}|\nu_{\alpha}(\lambda)\rangle,
	\label{Transition_ampl}
	\end{equation}
	satisfies
	\begin{equation}
	i\frac{d}{d\lambda}\Psi_{\alpha\beta}=\Phi(\lambda)\Psi_{\alpha\beta}.
	\label{Evol_Trans_Amp}
	\end{equation}
	The ultimate goal is to find the transition probability between flavors $\alpha$ and $\beta$, i.e.
	\begin{equation}
	P_{\beta\rightarrow\alpha}=|\Psi_{\alpha\beta}|^2=|\langle\nu_{\beta}|\nu_{\alpha}(\lambda)\rangle|^2,
	\label{Probability_Ini}
	\end{equation}
	and therefore we need to calculate $\Phi(\lambda)$, or more specifically, $P_{\mu}dx^{\mu}/d\lambda$, as has been pointed out before~\cite{Cardall-Fuller,Khalifeh:2020bdg}. 
	
	For the purpose at hand, let us start with a system of two-flavors, electron and muon neutrinos $\nu_e$ and $\nu_{\mu}$, respectively, that is $\alpha,\beta=e,\mu$. Let
	\begin{align}
	\boldsymbol{\psi} &= \begin{pmatrix}
	\psi_e\\
	\psi_{\mu}
	\end{pmatrix},
	\end{align}
	be a vector of spinor fields. The modified Dirac equation~\eqref{E.of.M} for this system becomes:
	\begin{equation}
	\bigg(i\gamma^{\mu}\mathcal{D}_{\mu}-\mathcal{M}_f\bigg)\boldsymbol{\psi}=\big(\xi F(\varphi,X_{\varphi}^{\mu})+\xi_f\gamma^{\mu}G_{\mu}(\varphi, X_{\varphi}^{\mu})\big)\boldsymbol{\psi}
	\label{New_Dirac}
	\end{equation}
	where the vacuum mass matrix in flavor space is given by
	\begin{align}
	\mathcal{M}_f^2 &= U\begin{pmatrix}
	m_1^2 & 0 \\
	0 & m_2^2
	\end{pmatrix}U^{\dagger}
	\label{eq:MassMatrix}
	\end{align}
	with $m_1, m_2$ being the masses of mass states $|\nu_1\rangle, |\nu_2\rangle$, respectively.
	
	Now we introduce the explicit form of the covariant derivative $\mathcal{D}_{\mu}$. As explained in Ref.~\cite{Birrell:1982ix,Lanzagorta}, when studying spinors in curved spacetime, one needs to introduce a local inertial coordinate system, with its own set of Dirac matrices $\gamma^a$, and link it to the general one using tetrad fields $e^{\mu}_a$, where Greek indices correspond to general coordinates, while Latin ones for the local system. With this, we can write
	\begin{equation}
	\gamma^{\mu}\mathcal{D}_{\mu}=\gamma^ae^{\mu}_a\big(\partial_{\mu}+\Gamma_{\mu}\big)
	\label{Cov_Deriv}
	\end{equation} 
	where
	\begin{equation}
	\Gamma_{\mu}=\frac{1}{8}\big[\gamma^b,\gamma^c\big]e_b^{\nu}\nabla_{\mu}e_{c\nu}
	\label{Spin_Conn}
	\end{equation}
	is called the spin-connection which takes into account the gravitational effect on the particle's spin. In eq.~\eqref{Spin_Conn}, $[\gamma^b,\gamma^c]$ is the commutator of $\gamma^a$ and $\gamma^b$, and $\nabla_{\mu}$ is the usual covariant derivative of General Relativity~\cite{Carroll:2004st}. With these relations, one can then show that
	\begin{equation}
	\gamma^ae_a^{\mu}\Gamma_{\mu}=i\gamma^ae_a^{\mu}A_{G\mu}
	\label{gamma_e_Gamma}
	\end{equation} 
	with
	\begin{equation}
	A_{G}^{\mu}=\frac{1}{4}\sqrt{-g}e_a^{\mu}\epsilon^{abcd}(\partial_{\sigma}e_{b\nu}-\partial_{\nu}e_{b\sigma})e_c^{\nu}e_d^{\sigma}
	\label{A_G}
	\end{equation}
	where $\epsilon^{abcd}$ is the local four dimensional Levi-Civita symbol. Inserting eqs.~\eqref{Cov_Deriv},~\eqref{gamma_e_Gamma} and~\eqref{A_G} in eq.~\eqref{New_Dirac} and moving all terms to the l.h.s, we get:
	\begin{equation}
	\bigg\{i\gamma^{\mu}\bigg[\partial_{\mu}+i\bigg(A_{G\mu}+\xi_fG_{\mu}\bigg)\bigg]-\bigg[\mathcal{M}_f-\xi F\bigg]\bigg\}\boldsymbol{\psi}=0.
	\label{Final_Mod_Dirac}
	\end{equation}
	In order to get non-trivial solutions for the above system, the determinant of the braces must be 0. This results in a modified mass-shell relation:
	\begin{equation}
	\bigg(P^{\mu}+A^{\mu}\bigg)\bigg(P_{\mu}+A_{\mu}\bigg)=\tilde{\mathcal{M}}_f^2,
	\label{Mod_Mass_Shell}
	\end{equation}
	where $A^{\mu}=A_G^{\mu}+\xi_fG_{\mu}$ and
	\begin{align}
	\tilde{\mathcal{M}}_f^2 &= U\begin{pmatrix}
	\tilde{m}_1^2 & 0 \\
	0 & \tilde{m}_2^2
	\end{pmatrix}U^{\dagger}
	\end{align} 
	with $\tilde{m}_i=m_i-\xi F$ for $i=1,2$. It should be noted here that $\tilde{m}$ is not a mass-varying neutrino, rather an effective mass due to the interaction with another field (see~\cite{PhysRevD.73.083515,Fardon_2004,Kaplan-Nu} for comparison). From eq.~\eqref{Mod_Mass_Shell}, one can show that
	\begin{equation}
	P_{\mu}\frac{dx^{\mu}}{d\lambda}=\frac{1}{2}\tilde{\mathcal{M}}_f^2-\frac{dx^{\mu}}{d\lambda}A_{\mu},
	\label{P_q_Prod}
	\end{equation}
	which finally implies, from eqs.~\eqref{Evolution_Operator} and~\eqref{Evol_Trans_Amp}, that
	\begin{equation}
	i\frac{d}{d\lambda}\Psi_{\alpha\beta}=\bigg[\frac{1}{2}\tilde{\mathcal{M}}_f^2+V_I\bigg]\Psi_{\alpha\beta},
	\label{Evol_Trans_Amp_2}
	\end{equation}
	with $V_I=-A_{\mu}dx^{\mu}/d\lambda$. In deriving eq.~\eqref{P_q_Prod}, two well motivated assumptions have been made based on the fact that we are focusing on high energy neutrinos~\cite{Cardall-Fuller,Khalifeh:2020bdg}. First, we consider neutrinos as energy eigenstates, i.e. $P^0=dx^0/d\lambda$, and second, $P^{i}$ and $dx^{i}/d\lambda$ are assumed parallel, that is $P^i=(1-\varepsilon)dx^i/d\lambda$, with $\varepsilon\ll1$ for high-energy neutrinos\footnote{The second condition can be relaxed since we are eventually taking the inner product of the two vectors, so that the perpendicular part does not contribute.}.
	
	\subsection{Transition Probability}
	\label{Trans_Prob}
	Let us now be more explicit, and look at each component of eq.~\eqref{Evol_Trans_Amp_2}. With some matrix algebra, it can be shown that
	\begin{align}
	\tilde{\mathcal{M}}_f^2 &=\bigg(\tilde{m}_1^2+\frac{1}{2}\tilde{\Delta}\bigg)I +\frac{1}{2}\tilde{\Delta}\begin{pmatrix}
	-\cos 2\theta & \sin 2\theta \\
	\sin 2\theta & \cos 2\theta
	\end{pmatrix},
	\label{Mod_Mass_Matrix}
	\end{align}
	where $I$ is the $2\times2$ identity matrix in flavor space and $\tilde{\Delta}_m^2=\tilde{m}_2^2-\tilde{m}_1^2=\Delta_m^2-2\xi F\Delta_m$(up to 1${}^{\text{st}}$ order in $\xi$), with $\Delta_m^2=m_2^2-m_1^2$ and $\Delta_m=(m_2-m_1)$. Note the resemblance to the MSW effect~\cite{Mikheev:1986gs,MSW2}, with the difference being that interactions with matter are substituted by those with spacetime and DE. Also, it is safe to ignore the term proportional to $I$ in eq.~\eqref{Mod_Mass_Matrix}, since it is common for both transition amplitudes, and therefore will cancel when we calculate the probability. If we start initially from a $\nu_e$ state, for instance, the evolution equation for the transition amplitudes becomes:
	\begin{align}
	i\frac{d}{d\lambda}\begin{pmatrix}
	\Psi_{ee} \\
	\Psi_{e\mu}
	\end{pmatrix} &=\begin{pmatrix}
	-\frac{1}{4}\tilde{\Delta}_m^2\cos2\theta+\xi_eV_I & \frac{1}{4}\tilde{\Delta}_m^2\sin2\theta\\ 
	\frac{1}{4}\tilde{\Delta}_m^2\sin2\theta & \frac{1}{4}\tilde{\Delta}_m^2\cos2\theta+\xi_{\mu}V_I
	\end{pmatrix}
	\begin{pmatrix}
	\Psi_{ee} \\
	\Psi_{e\mu}
	\end{pmatrix}\nonumber\\
	&\equiv \boldsymbol{M}\begin{pmatrix}
	\Psi_{ee} \\
	\Psi_{e\mu}
	\end{pmatrix}.
	\label{Evol_Eq_Fin}
	\end{align}
	Notice that the gravitational contribution $A_{G\mu}$ has been dropped from the interaction term. This is because it is proportional to $I$ in flavor space, and therefore does not contribute to the oscillation probability~\cite{Cardall-Fuller}. In addition to that, in spatially homogeneous and isotropic universes, such as FRW, this term is 0 identically~\cite{Khalifeh:2020bdg}.
	
	From eq.~\eqref{Evol_Eq_Fin}, we can proceed by diagonalizing $\boldsymbol{M}$, which has
	\begin{equation}
	v_{\pm}=\frac{1}{4}\bigg[2\big(\xi_e+\xi_{\mu}\big)V_I\pm\sqrt{\big[\tilde{\Delta}_m^2\cos2\theta-2V_I(\xi_e-\xi_{\mu})\big]^2+(\tilde{\Delta}_m^2)^2\sin^22\theta}\bigg]
	\label{Eigenvalues_Utild}
	\end{equation}
	as  eigenvalues, and
	\begin{align}
	\tilde{U} &=\begin{pmatrix}
	\cos\tilde{\theta} & \sin\tilde{\theta}\\
	-\sin\tilde{\theta} & \cos\tilde{\theta}
	\end{pmatrix}
	\label{Utild}
	\end{align}
	as the unitary matrix that diagonalizes it, with\footnote{One way of deriving eq.~\eqref{Sin_Cos_tild} is to perform the matrix product $\tilde{U}^T\boldsymbol{M}\tilde{U}$, and equate it to $\text{diag}\{v_-,v_+\}$.}
	\begin{align}
	\cos 2\tilde{\theta}=\frac{\tilde{\Delta}_m^2\cos 2\theta-2V_I\big(\xi_e-\xi_{\mu}\big)}{\sqrt{\big[\tilde{\Delta}_m^2\cos 2\theta-2V_I\big(\xi_e-\xi_{\mu}\big)\big]^2+(\tilde{\Delta}_m^2)^2\sin^22\theta}};\nonumber\\
	\sin 2\tilde{\theta}=\frac{\tilde{\Delta}_m^2\sin 2\theta}{\sqrt{\big[\tilde{\Delta}_m^2\cos 2\theta-2V_I\big(\xi_e-\xi_{\mu}\big)\big]^2+(\tilde{\Delta}_m^2)^2\sin^22\theta}}.
	\label{Sin_Cos_tild}
	\end{align}
	In analogy with the flavor-mass bases transformation, let us define
	\begin{align}
	\boldsymbol{\phi}_e &\equiv \begin{pmatrix}
	\phi_{e-} \\
	\phi_{e+}
	\end{pmatrix}= \tilde{U}^T\begin{pmatrix}
	\Psi_{ee} \\
	\Psi_{e\mu}
	\end{pmatrix},
	\label{Phi}
	\end{align}
	as a vector of transition amplitudes in an effective mass basis, $\{\nu_-,\nu_+\}$, that takes into account the neutrino interaction with gravity and DE. Using the unitarity of $\tilde{U}$ and eq.~\eqref{Evol_Eq_Fin}, it can be shown that $\boldsymbol{\phi}_e$ satisfies:
	\begin{align}
	i\frac{d}{d\lambda}\begin{pmatrix}
	\phi_{e-} \\
	\phi_{e+}
	\end{pmatrix}=\begin{pmatrix}
	v_- & -i\frac{d\tilde{\theta}}{d\lambda}\\
	i\frac{d\tilde{\theta}}{d\lambda} & v_+
	\end{pmatrix}
	\begin{pmatrix}
	\phi_{e-} \\
	\phi_{e+}
	\end{pmatrix}.
	\end{align}
	Notice that the off-diagonal terms come from transforming the l.h.s of eq.~\eqref{Evol_Eq_Fin}. Also, in the case where there is no mixing between effective mass states, then $d\tilde{\theta}/d\lambda=0$, and the transition amplitudes evolve as:
	\begin{equation}
	\phi_{ej}=\big(\cos\omega_j+i\sin\omega_j\big)\phi_{ej}(0)
	\end{equation}
	for $j=+,-$, where $\phi_{ej}(0)$ is the initial condition and
	\begin{equation}
	\omega_j(\lambda)=\int_{\lambda_0}^{\lambda}v_jd\lambda'.
	\end{equation}
	This is known as the adiabatic evolution condition which, as we will see in the next section, applies to the DE scenarios we will examine. Further, one important consequence of adiabaticity is that the flavor-specific interaction will be constant along the neutrino's world-line. To see this, differentiate $\cos 2\tilde{\theta}$ from eq.~\eqref{Sin_Cos_tild} w.r.t $\lambda$:
	\begin{equation}
	\frac{d\tilde{\theta}}{d\lambda}=\frac{\sin 2\tilde{\theta}}{\Delta_m}\biggl\{-2\xi\Delta_m\frac{dF}{d\lambda}\bigg[\frac{\cos 2\tilde{\theta}\sin 2\theta}{\sin 2\tilde{\theta}}-\cos 2\theta\bigg]
	+\frac{dV_I}{d\lambda}(\xi_e-\xi_{\mu})\biggr\}.
	\label{Tild_Theta_Evol}
	\end{equation}
	Since we expect gravitational and DE effects to be small compared to the vacuum oscillations, that is $\xi,\xi_f\ll1$, we can keep terms up to first order in these coupling constants. With this assumption, by setting eq.~\eqref{Tild_Theta_Evol} to 0, we find that $dV_I/d\lambda=0$. This is again another analogy with the MSW effect, where in adiabatic oscillations the interaction term is constant along the path~\cite{Giunti:2007ry}.
	
	The final ingredient we need to get the oscillation probability is initial conditions. As we are considering an initial $\nu_e$ state, we can write
	\begin{align}
	\boldsymbol{\phi}_e(0) &\equiv \begin{pmatrix}
	\phi_{e-}(0) \\
	\phi_{e+}(0)
	\end{pmatrix}=\begin{pmatrix}
	\cos\tilde{\theta} & -\sin\tilde{\theta}\\
	\sin\tilde{\theta} & \cos\tilde{\theta}
	\end{pmatrix}\begin{pmatrix}
	1 \\ 0
	\end{pmatrix}=\begin{pmatrix}
	\cos\tilde{\theta} \\ 
	\sin\tilde{\theta}
	\end{pmatrix},
	\label{Ini_Con_Phi}
	\end{align}
	where, by construction, an initial $\nu_e$ state corresponds to $\Psi_{ee}(0)=1$ and $\Psi_{e\mu}(0)=0$. By acting with the inverse transformation of eq.~\eqref{Phi}, we can calculate the amplitude $\Psi_{e\mu}$, and thus, with the initial conditions eq.~\eqref{Ini_Con_Phi}, we finally obtain the $\nu_e\rightarrow\nu_{\mu}$ transition probability:
	\begin{equation}
	P_{\nu_e\rightarrow\nu_{\mu}}=|\Psi_{e\mu}|^2=\sin^2 2\tilde{\theta}\sin^2\bigg(\frac{\omega_--\omega_+}{2}\bigg).
	\label{Probability}
	\end{equation}
	
	Let us now look in more detail into the oscillating term in eq.~\eqref{Probability}. If we use the above mentioned approximation ($\xi$, $\xi_f\ll1$), one can show that
	\begin{align}
	\omega_--\omega_+ &=\int_{\lambda_0}^{\lambda}(v_--v_+)d\lambda'\nonumber
	\\
	&\approx\frac{\Delta_m^2}{2}(\lambda_0-\lambda)+V_I\cos 2\theta(\xi_e-\xi_{\mu})(\lambda-\lambda_0)+\xi\Delta_m\int_{\lambda_0}^{\lambda}Fd\lambda'.
	\label{OmegaMinusOmega}
	\end{align}
	The first term in eq.~\eqref{OmegaMinusOmega} corresponds to the usual vacuum oscillation term. Indeed, if one neglects the interactions completely, i.e. $F=G=0$, and consider Minkowski spacetime, that is $d\lambda=dt/E= dx/E$, we get
	\begin{equation}
	\omega_--\omega_+=\omega_{\text{std}}\equiv\frac{\Delta_m^2L}{2E}
	\end{equation}
	where $L$ is the distance traveled by the neutrinos. This results in
	\begin{equation}
	P^{\text{std}}_{\nu_e\rightarrow\nu_{\mu}}=\sin^2 2\theta\sin^2\bigg(\frac{\Delta_m^2L}{4E}\bigg),
	\label{Std_Prob}
	\end{equation}
	which is the standard vacuum transition probability in flat spacetime~\cite{Giunti:2007ry}. The second term in eq.~\eqref{OmegaMinusOmega} is the flavor-specific correction, and the third term is an integrated correction from the flavor-invariant interaction. This is where we see that the latter does affect the transition probability, both in amplitude, through $\sin 2\tilde{\theta}$ in eq.~\eqref{Probability},  and in period.
	
	Let us finish the analysis by writing a Signal-to-Noise-like expression for the oscillation probability eq.~\eqref{Probability}. If we substitute the expression for $\sin 2\tilde{\theta}$ from eq.~\eqref{Sin_Cos_tild} into eq.~\eqref{Probability}, and then expand all functions of the interactions up to 1${}^{\text{st}}$ order, we get
	\begin{equation}
	\frac{\delta P}{P}\equiv \frac{P_{\nu_e\rightarrow\nu_{\mu}}-P^{\text{vac}}_{\nu_e\rightarrow\nu_{\mu}}}{P^{\text{vac}}_{\nu_e\rightarrow\nu_{\mu}}}=\frac{4V_I(\xi_e-\xi_{\mu})}{\Delta_m^2}+\cot\bigg(\frac{1}{2}\omega_{\text{vac}}\bigg)\omega_{\text{DE}},
	\label{Signal_to_noise}
	\end{equation} 
	where
	\begin{equation}
	P^{\text{vac}}_{\nu_e\rightarrow\nu_{\mu}}=\sin^22\theta\sin^2\omega_{\text{vac}}
	\end{equation}
	is the transition probability in vacuum, with frequency
	\begin{equation}
	\omega_{\text{vac}}=\frac{\Delta_m^2}{2}(\lambda_0-\lambda),
	\end{equation}
	and
	\begin{equation}
	\omega_{\text{DE}}= V_I\cos 2\theta(\xi_e-\xi_{\mu})(\lambda-\lambda_0)+\xi\Delta_m\int_{\lambda_0}^{\lambda}Fd\lambda'
	\end{equation}
	is the additional contribution to the oscillation frequency due to the interaction with DE.

	In this section, we looked at how a type of general interactions between neutrinos and DE, in a generic spacetime, can affect the probability of oscillations, with the final result given in eq.~\eqref{Probability}. Now we can specify the interaction to known DE models, particularly a cosmological constant and scalar field based DE, and thus establish the distinction between them.
	
	
	\section{Oscillation Probability for Specific DE Models}
	
	As mentioned in the Introduction~\ref{Intro.}, we have focused on the interaction of neutrinos with scalar fields since the latter includes a large class of DE models, such as some modified gravity scenarios and Quintessence. Having established a general formalism for the interaction of neutrinos with a scalar field in the previous section, we will now focus on two DE energy models: a Cosmological Constant $\Lambda$ and Quintessence.
	\subsection{$\Lambda$CDM}
	\label{Sec:LCDM}
	This model is the simplest model describing our Universe, and has sustained a great deal of observational test~\cite{Planck2018,BOSS}. Taking GR as the theory of gravity, $\Lambda$CDM has two main components in the late universe: a cosmological constant DE, $\Lambda$, and cold Dark Matter(CDM). The metric of spacetime that best describes it is FLRW:
	\begin{equation}
	ds^2=-dt^2+a^2(t)\delta_{ij}dx^idx^j,
	\end{equation}
	where $t$ is cosmic time, $x^i$, for $i=1,2,3$, are comoving spatial coordinates, $\delta_{ij}$ is the Kronecker delta and $a(t)$ is the scale factor that incorporates the universe's expansion. The resulting evolution equation for the scale factor will be the usual first Friedmann equation:
	\begin{equation}
	H^2=H_0^2\big(\Omega_{m_0}a^{-3}+\Omega_{\Lambda_0}\big),
	\label{Friedmann}
	\end{equation}
	where $H=\dot{a}/a$ is the Hubble parameter, with $H_0$ being its value today and $\Omega_{m0}$, $\Omega_{\Lambda 0}$ are today's matter and $\Lambda$ density parameters, respectively, with the most recent measurement given by the Planck Collaboration~\cite{Planck2018}. Note that this equation is not altered for two reasons. First, we are considering gravity in terms of the background spacetime, and not from a quantum perspective, hence Einstein equations, from which eq.~\eqref{Friedmann} is derived, are still the same. Second, the neutrino density parameter has not been added since it is small compared to that of matter and $\Lambda$~\cite{Dodelson}.
	
	To see the effect of DE in this model on neutrino oscillations, let us start by noting that when DE is $\Lambda$, $\xi=\xi_f=0$ in eq.~\eqref{Gen.Inter.}, implying the automatic satisfaction of the adiabaticity condition, $d\tilde{\theta}/d\lambda=0$, from eq.~\eqref{Tild_Theta_Evol}, and thus $\sin\tilde{\theta}(\cos\tilde{\theta})=\sin\theta(\cos\theta)$. Therefore, from the first equality in eq.~\eqref{OmegaMinusOmega}, we define
	\begin{equation}
	\omega_{\Lambda}\equiv\omega_--\omega_+\big|_{DE=\Lambda}=\frac{\Delta_m^2}{2}\int_{t_{\text{em}}}^{t_0}\frac{1}{E}dt,
	\label{Omega_minus_Omega_Lambda}
	\end{equation}
	with the second equality meaning eq.~\eqref{OmegaMinusOmega} when DE is $\Lambda$. Here, $t_0$ is today, $t_{\text{em}}$ is the time of neutrino emission and $E=dt/d\lambda$ is the 0${}^{\text{th}}$ component of the null tangent vector $dx^{\mu}/d\lambda$, which is also the neutrino's energy. Since the latter follows the geodesic equation, as shown in~\cite{Cardall-Fuller,Khalifeh:2020bdg}, then $E=E_0/a$, with $E_0$ being the neutrino energy at detection, and thus, using eq.~\eqref{Friedmann},
	\begin{align}
	\omega_{\Lambda}& =\frac{\Delta_m^2}{2H_0E_0}\int_{a_{\text{em}}}^{1}\bigg(\Omega_{m_0}a^{-3}+\Omega_{\Lambda_0}\bigg)^{-1/2}da\nonumber \\
	&=\frac{\Delta_m^2}{2H_0E_0}\int_{0}^{z_{\text{em}}}\bigg(\Omega_{m_0}(1+z)^7+\Omega_{\Lambda_0}(1+z)^4\bigg)^{-1/2}dz,
	\label{omega_Lambda}
	\end{align}
	where $a_{\text{em}}(z_{\text{em}})$ is the scale factor(redshift) at neutrino emission, and with the usual normalization $a_0=1$and $z_0=0$. On the other hand, if one takes a simple approach (SA) to neutrino oscillations in an expanding universe, and substitutes $L$ and $E$ in eq.~\eqref{Std_Prob} by the luminosity distance,
	\begin{equation}
	D_L=(1+z_e)H_0^{-1}\int_0^{z_e}\bigg(\Omega_{m_0}(1+z)^{3}+\Omega_{\Lambda_0}\bigg)^{-1/2}dz
	\label{Dl}
	\end{equation}
	and $E=E_0(1+z_e)$, respectively, we get,
	\begin{equation}
	\omega_{\text{SA}}=\frac{\Delta_m^2}{2H_0E_0}\int_0^{z_e}\bigg(\Omega_{m_0}(1+z)^{3}+\Omega_{\Lambda_0}\bigg)^{-1/2}dz.
	\label{omega_Dl}
	\end{equation}
	Finally, inserting eqs.~\eqref{omega_Lambda} and~\eqref{omega_Dl} in eq.~\eqref{Probability} gives the two-flavor neutrino oscillation probability in the $\Lambda$CDM model,
	\begin{equation}
	P_{\Lambda}=\sin^22\theta\sin^2\omega_{\Lambda},
	\label{P_Lambda}
	\end{equation}
	and in the SA,
	\begin{equation}
	P_{\text{SA}}=\sin^22\theta\sin^2\omega_{\text{SA}}.
	\label{P_SA}
	\end{equation}
	
\begin{figure}
			\centering
			\includegraphics[width=10cm]{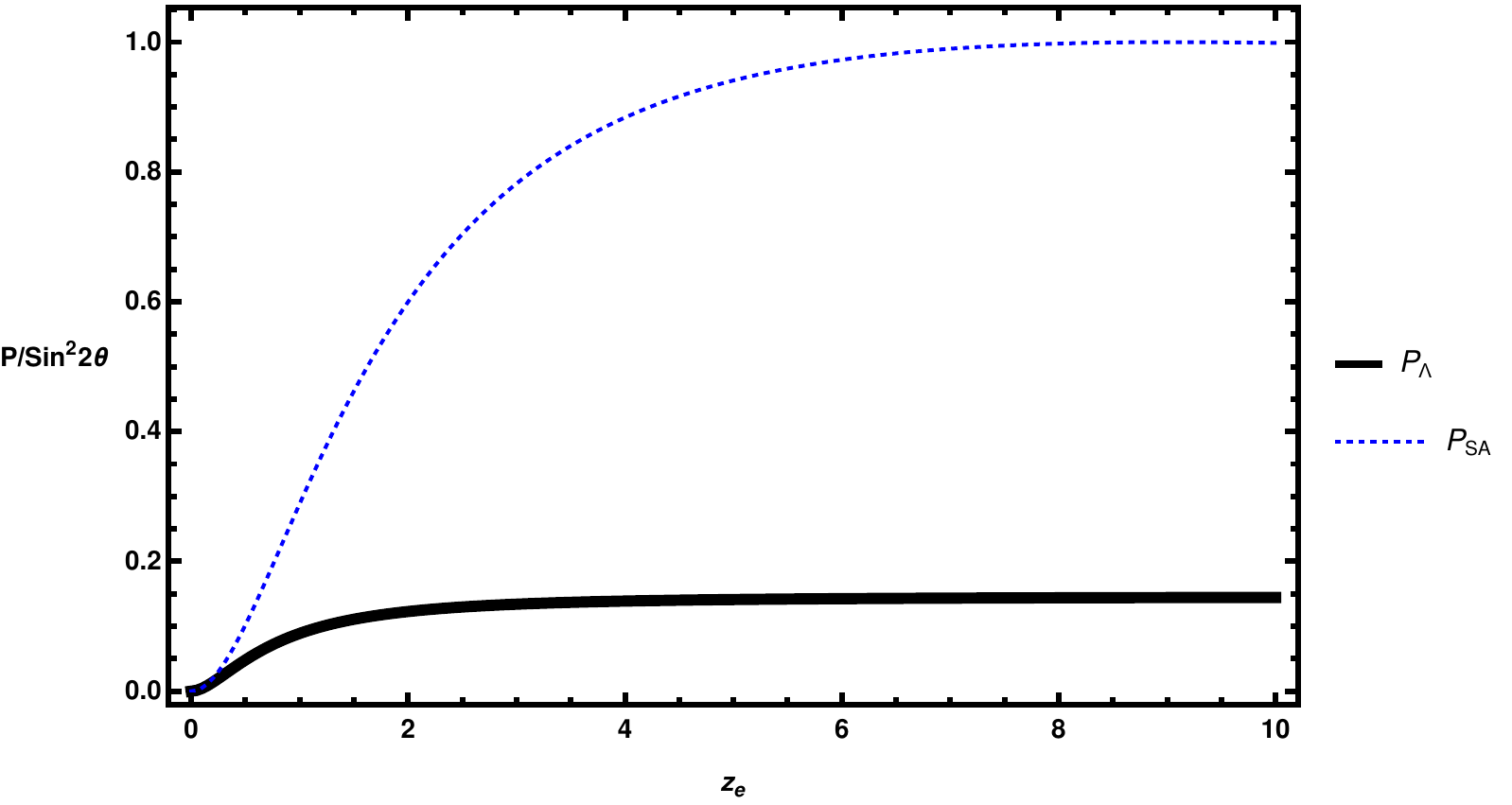}		
 			\caption{The two-flavor neutrino oscillation probability, divided by $\sin^22\theta$, as a function of redshift of emission, $z_e\in[0,10]$, for the two cases $P_{\Lambda}$(solid black curve) and $P_{\text{SA}}$(dotted blue line), given by eqs.~\eqref{Omega_minus_Omega_Lambda},\eqref{P_Lambda} and eqs.~\eqref{omega_Dl},\eqref{P_SA}, respectively. To be specific, due to the large value of $C=\Delta_m^2/(2H_0E_0)$, we used $\omega\mod{2\pi C}$ as the argument of $\sin^2$ in eqs.\eqref{P_Lambda}-\eqref{P_SA} to avoid numerical instabilities. The values of the different parameters used in these equations is given in the text below eq.\eqref{P_SA}.}
 			\label{fig:Plot_P_L_Dl}
	\end{figure}

\begin{figure}
	\centering
	\includegraphics[width=10cm]{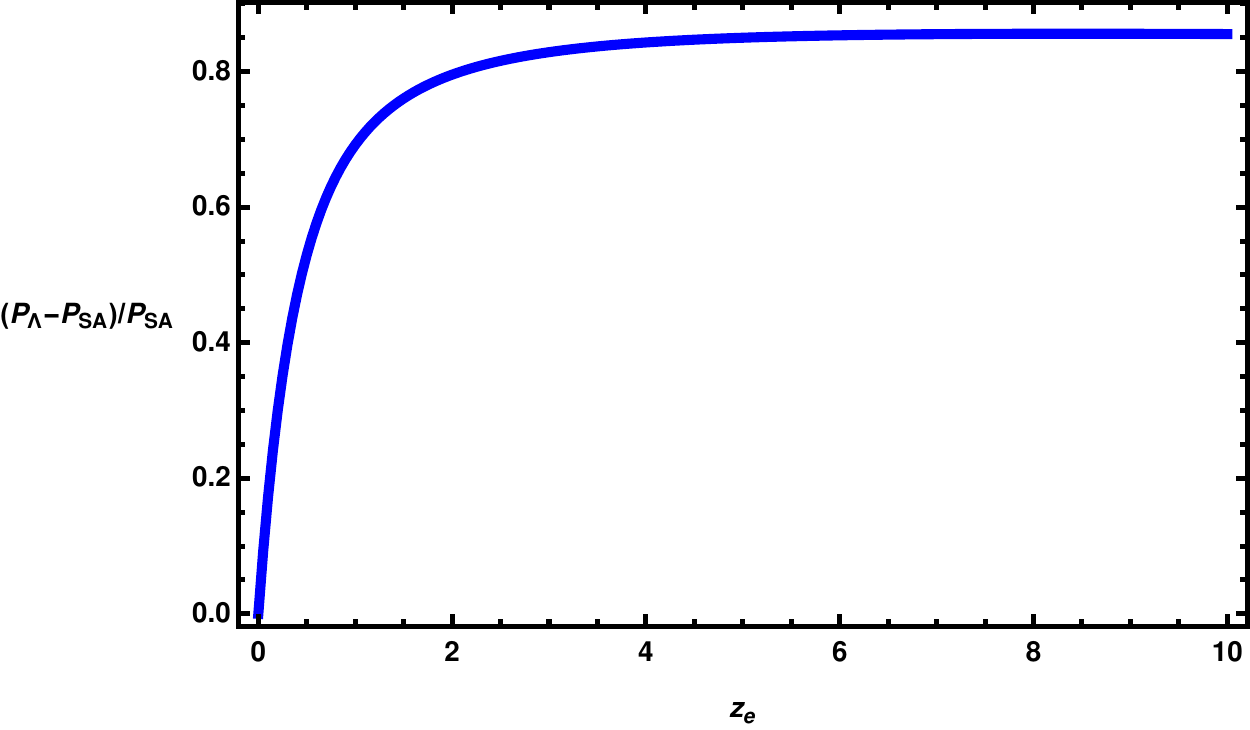}
	\caption{Absolute value of the fractional difference between the oscillation probabilities $P_{\Lambda}$ and $P_{\text{SA}}$, given by eqs.~\eqref{Omega_minus_Omega_Lambda},\eqref{P_Lambda} and eqs.~\eqref{omega_Dl},\eqref{P_SA}, respectively, as a function of emission redshift, $z_e\in[0,10]$. To be specific, due to the large value of $C=\Delta_m^2/(2H_0E_0)$, we used $\omega\mod{2\pi C}$ as the argument of $\sin^2$ in eqs.\eqref{P_Lambda}-\eqref{P_SA} to avoid numerical instabilities. Moreover, we shifted both $P_{\Lambda}$ and $P_{\text{SA}}$ by $10^{-5}$ to avoid the singularity when they are 0 at $z_e=0$. The values of different parameters used in these equations is given in the text below eq.\eqref{P_SA}.}
	\label{fig:Plot_diff_L_Dl}
	\end{figure}

	In Figure~\ref{fig:Plot_P_L_Dl}, we plot the evolution of $P_{\Lambda}$(solid black curve) and $P_{\text{SA}}$(dotted blue curve) as a function of redshift, to compare the two approaches. To this end, we took $\Delta_m^2=7.53\times 10^{-5}$eV${}^2$\footnote{Here we used mass states 1 and 2 from Ref.~\cite{Review_Particle_Physics} as ours. One can check that other values of $\Delta_m^2$ reported there does not alter the evolution of the frequencies eqs.(\ref{omega_Lambda}, \ref{omega_Dl}). Physically, this is due to the absence of a direct interaction between DE and neutrinos. Mathematically, this is because the coefficient multiplying the integrals in eqs.(\ref{omega_Lambda}, \ref{omega_Dl}) includes $H_0^{-1}\sim\mathcal{O}(10^{33})$eV, which wipes out the $\mathcal{O}(10^2)$eV${}^2$ difference between $\Delta_m^2$s.} and $E_0=10^{16}$eV, a value to which neutrino detectors are on average sensitive to~\cite{IceCube}. Further, we used $\Omega_{m_0}=0.315, \Omega_{\Lambda_0}=0.685$ and $H_0=1.44\times10^{-33}$eV as reported in~\cite{Planck2018}. For redshifts higher than $\sim2$, the difference between the two probabilities stabilizes at around 80\%, as can be seen from figure~\ref{fig:Plot_diff_L_Dl}. On the other hand, the latter shows, for the observationally more interesting range of redshifts ($0\leq z\leq 0.5$), the difference can reach up to 50\% while they coincide for redshift 0, as expected.
	
	The difference between eqs.(\ref{omega_Lambda}, \ref{P_Lambda}) and eqs.(\ref{omega_Dl},\ref{P_SA}) is being highlighted here to insure that, when doing neutrino observations, one cannot directly substitute $D_L(z)$ and $E(z)$ as neutrino traveling length and energy, respectively. This will not properly take into account the evolution of a spin 1/2 particle in a curved background. Rather, one should use the formalism presented in section~\ref{Nu_Oscill}, for a more general interaction with a scalar field in curved spacetime, or eqs.(\ref{omega_Lambda}, \ref{P_Lambda}) for $\Lambda$CDM. The same idea applies to other models of DE, however there will be differences in the evolution of the oscillation probability, as we will see next.


	\subsection{Quintessence}
	\label{Sec:Quintessence}
	
	As a homogeneous canonical scalar field minimally coupled to gravity, Quint-essence could be an explanation to the late time accelerated expansion~\cite{Peebled_Bharat,Wetterich:1987fm,Fujii_Yasunori,Ford_LH,Wang:Coincidence}. One of the main reasons for introducing quintessence as an alternative to a cosmological constant is to make DE dynamical, thereby avoiding the cosmological constant and coincidence problems(see~\cite{Tsujikawa:2013fta} and references therein for more information on Quintessence).
	
	In order to probe this model using neutrino oscillations, a coupling between the scalar and spinor fields has to be introduced, otherwise the difference in effect of quintessence and $\Lambda$ on the oscillation probability will be difficult to observe. We consider the coupling introduced in~\cite{2018PDU....20...72S}, given the name DE${}_{\nu}$ model, and which we analyzed in~\cite{Khalifeh:2020bdg}. In the present formalism, DE${}_{\nu}$ translates to $F=0$ and $G_{\mu}=\partial_{\mu}\varphi$ in eq.~\eqref{Gen.Inter.}. As mentioned in~\cite{2018PDU....20...72S}, such a derivative coupling is a low energy limit of the model presented in~\cite{Fukugita}, with the scalar field being a Nambu-Goldstone boson resulting from the spontaneous symmetry breaking of Lepton number symmetry~\cite{burgess:effectivelagrangian,Witten:2000dt,Mohapatra:1980ui}. This shows that such a coupling is motivated both from Particle Physics and Cosmology points of view, hence it is being further analyzed here.
	
	To start the analysis, recall that since the scalar field is homogeneous, its energy density would be
	\begin{equation}
	\rho_{\varphi}=\frac{1}{2}\dot{\varphi}^2+V(\varphi)
	\end{equation}
	where $\dot{\varphi}=\partial_t\varphi$ and $V_{\varphi}(\varphi)$ is the potential energy of $\varphi$. Therefore, eq.~\eqref{Friedmann} becomes
	\begin{equation}
	H^2=H_0^2\big(\Omega_{m_0}a^{-3}+\Omega_{\varphi}\big),
	\label{Friedmann_Quint}
	\end{equation}
	where
	\begin{equation}
	\Omega_{\varphi}=\frac{8\pi G}{3H_0^2}\rho_{\varphi}
	\label{Omega_Quint}
	\end{equation}
	is the density parameter of quintessence. Moreover, as already shown in~\cite{Khalifeh:2020bdg}, this type of interactions does not affect the Klein-Gordon equation, which can be written as:
	\begin{equation}
	\frac{d}{da}\big(a^6\dot{\varphi}^2\big)=2a^6\frac{dV}{da}.
	\label{Klein-Gordon}
	\end{equation}
	For $\varphi$ to produce an accelerated expansion, it should satisfy the condition:
	\begin{equation}
	\dot{\varphi}^2\ll V(\varphi)\approx\rho_{\Lambda_0},
	\label{Accel_Cond}
	\end{equation}
	where $\rho_{\Lambda_0}$ is the energy density of a cosmological constant today. This means that, first, in eq.~\eqref{Friedmann_Quint}, $\Omega_{\varphi}\approx\Omega_{\Lambda_0}$, and second, we can write\footnote{The $7/2$ factor is to reduce numerical factors clustering. }
	\begin{equation}
	\frac{dV}{da}\equiv\frac{7}{2}\epsilon,
	\label{Potential}
	\end{equation}
	where $\epsilon<\rho_{\Lambda_0}\sim\mathcal{O}(10^{-11}\text{eV}^4)$\footnote{\label{eps} Note that $\epsilon$ is not exactly the slow-roll parameter $\varepsilon=d(H^{-1})/dt$, but one can show that $\varepsilon\approx3\epsilon a/V$.}, and thus, from eq.~\eqref{Klein-Gordon}, we get
	\begin{equation}
	\dot{\varphi}=\sqrt{\epsilon a}.
	\label{phi_dot}
	\end{equation}

	On the neutrino's side, this type of interaction results in
	\begin{equation}
	V_I=-\frac{dx^{\mu}}{d\lambda}G_{\mu}=-E\dot{\varphi}=-E_0\sqrt{\epsilon(1+z)},
	\label{V_Quint}
	\end{equation}
	from which one can show, using eqs.~\eqref{Eigenvalues_Utild},~\eqref{Sin_Cos_tild} and~\eqref{Tild_Theta_Evol}, that
	\begin{equation}
	v_{\pm}=\frac{1}{2}V_I\big[\xi_e+\xi_{\mu}\mp\cos2\theta(\xi_e-\xi_{\mu})\big]\pm\Delta_m^2,
	\label{v_pm_Quint}
	\end{equation}
	\begin{equation}
	\sin2\tilde{\theta}=\sin2\theta\bigg[1+\frac{4E_0\sqrt{\epsilon(1+z)}}{\Delta_m^2}\cos2\theta(\xi_e-\xi_{\mu})\bigg]^{-1/2}
	\label{Sin_tild_Quint}
	\end{equation}
	and
	\begin{equation}
	\frac{d\tilde{\theta}}{d\lambda}=\frac{\sin2\tilde{\theta}}{\Delta_m}(\xi_e-\xi_{\mu})\frac{dV_I}{d\lambda},
	\label{Tild_Theta_Evol_Q}
	\end{equation}
	respectively. To check if the adiabaticity condition is satisfied for the current case, differentiate eq.~\eqref{V_Quint} w.r.t $\lambda$ and insert it in eq.~\eqref{Tild_Theta_Evol_Q}, to find
	\begin{equation}
	\frac{d\tilde{\theta}}{d\lambda}\approx\frac{\sin2\theta(\xi_e-\xi_{\mu})}{2\Delta_m}E_0^2\epsilon^{1/2} H_0\sqrt{\Omega_{m_0}(1+z)^6+\Omega_{\varphi_0}(1+z)^3}.
	\end{equation}
	From the fact that $H_0\sim\mathcal{O}(10^{-33})$eV~\cite{Dodelson}, $E_0\sim\mathcal{O}(10^{16})$eV(typical value for high-energy neutrinos~\cite{IceCube}), $\xi_{e,\mu}\sim\mathcal{O}(10^{-14})$eV${}^{-1}$~\cite{Khalifeh:2020bdg} and $\epsilon\sim\mathcal{O}(10^{-11})$eV${}^{4}$, one can see that $d\tilde{\theta}/d\lambda\ll v_{\pm}$, and therefore the adiabaticity condition still holds, resulting in an oscillation frequency
	\begin{align}
	\omega_{Q}\approx& \frac{\Delta_m^2}{2E_0H_0}\int_{0}^{z_e}\sqrt{\frac{1+\frac{4E_0\sqrt{\epsilon(1+z)}}{\Delta_m^2}\cos2\theta(\xi_e-\xi_{\mu})}{\Omega_{m_0}(1+z)^{7}+\Omega_{\varphi_0}(1+z)^4}}dz.
	\end{align}
	Finally, from eq.~\eqref{Probability}, the two-flavor oscillation probability in the case when DE is quintessence is:
	\begin{equation}
	P_Q=\frac{\sin^22\theta}{1+\frac{4E_0\sqrt{\epsilon(1+z)}}{\Delta_m^2}\cos2\theta(\xi_e-\xi_{\mu})}\sin^2(\omega_Q/2).
	\label{P_Q}
	\end{equation}
	
\begin{figure}
		\centering
		\includegraphics[width=1.1\columnwidth]{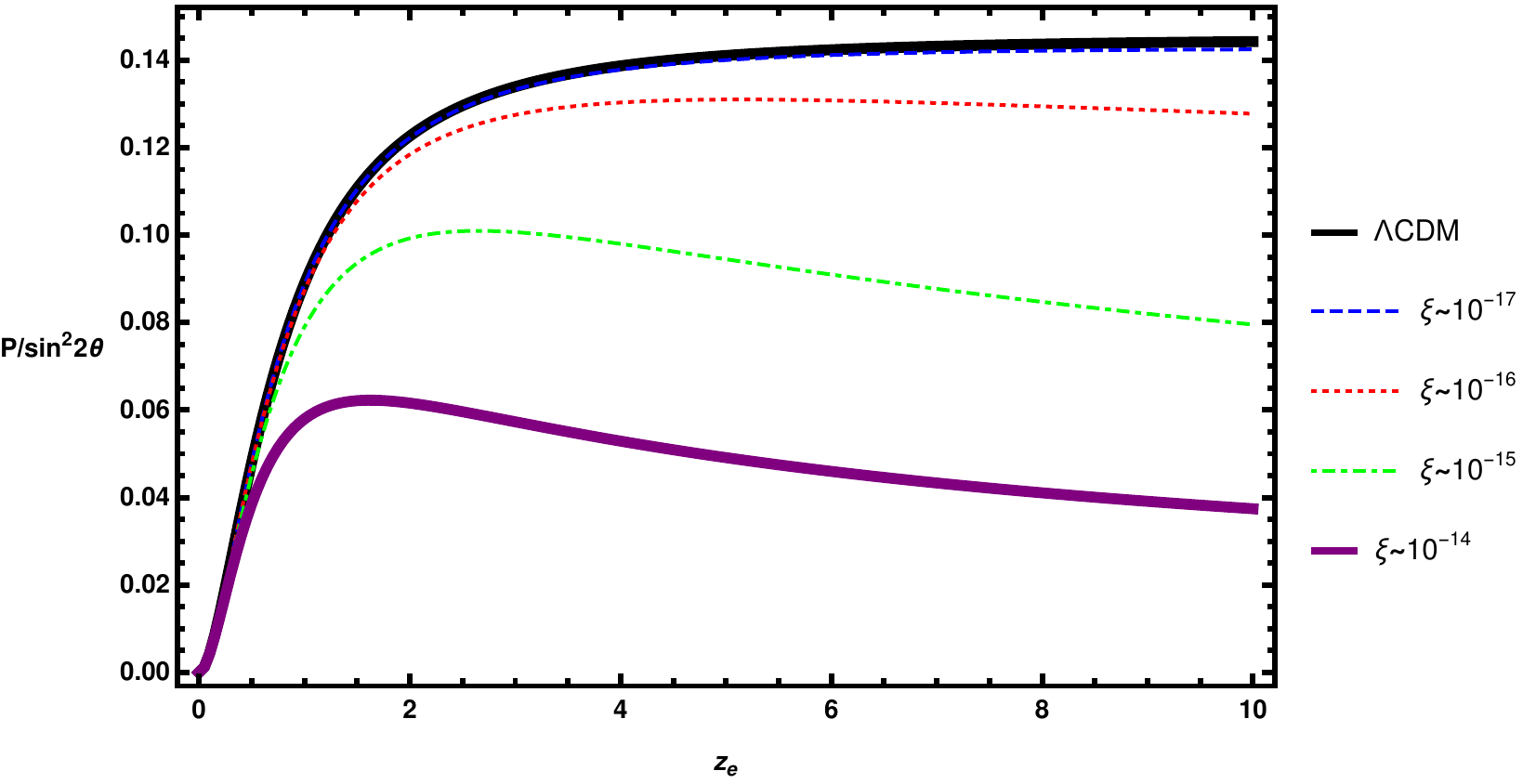}
		\caption{Evolution of the neutrino oscillation probability with redshift in the case of $\Lambda$CDM(solid black line) and quintessence, for neutrino-quintessence coupling $\sim\mathcal{O}(10^{-17})$(dashed blue), $10^{-16}$(dotted red), $10^{-15}$(dot-dash green) and $10^{-14}$(solid purple line). The parameters used are given in the text, after eq.~\eqref{P_SA}.}
		\label{fig:Plot_ProbQ_Chis}
	\end{figure}

	To study the difference between this model and $\Lambda$CDM, we plot(figure~\ref{fig:Plot_ProbQ_Chis}) eq.\eqref{P_Q} for values of $\xi_i$, $i=e,\mu$, ranging from $10^{-17}$ to $10^{-14}$eV${}^{-1}$, in addition to eq.\eqref{P_Lambda} for the $\Lambda$CDM case. We have checked that smaller values of $\xi_i$ do not produce any noticeable deviation from $\Lambda$CDM, while already at $10^{-14}$ we can see from figure~\ref{fig:Plot_ProbQ_Chis} that the deviation is $\sim50\%$ at $z=2$. That is the reason why we focus on this range of values of the couplings $\xi_i$. Moreover, we use the same parameters used to produce figures~\ref{fig:Plot_P_L_Dl} and~\ref{fig:Plot_diff_L_Dl}(see text after eq.~\eqref{P_SA}), in addition to $\cos2\theta=0.4$~\cite{Review_Particle_Physics}. As the strength of the coupling increases, the difference between the two models starts to become apparent at redshift $\sim0.5$, which is expected since then DE is becoming more dynamical than in the case of $\Lambda$CDM. 
	
	In order to make the distinction between the different DE scenarios more concrete, we study in more detail the dependence of $P_Q$ in eq.~\eqref{P_Q} on the parameters $\epsilon$ and $\xi_f$ of this particular DE model. First, if quintessence is slow-rolling, but not ultra slow-rolling, then $\varepsilon$(see footnote~\ref{eps}) cannot be too small~\cite{Inflation:JudgmentDay}. Taking $\epsilon\in[10^{-14},10^{-12}]$, which corresponds to $\varepsilon\in[10^{-3},10^{-1}]$, we find that $\xi_f\in[10^{-15}, 10^{-13}]$ gives distinguishable stable results. On the other hand, values beyond this interval would lead to unstable transition probabilities. This shows that, in principle, the presented method here could provide a complementary theoretical constraint to this type of coupling.
	
	Second, if we now fix $\epsilon\sim\mathcal{O}(10^{-13})$, for instance, we find that the difference between quintessence and $\Lambda$CDM starts to become appreciable(i.e. more that a few \%) for  $\xi_f\sim\mathcal{O}(10^{-14}-10^{-13})$. This is still consistent with Particle Physics constraints for this type of coupling, which is $\xi_f\lesssim10^{-7}$eV${}^{-1}$~\cite{Gando_2012,Kelly_2018}. Moreover, for values of $\xi_f$ that differ from each other by at least half an order of magnitude, the transition probabilities start deviating from each other by more than a few \%. One may conclude from this that there is a small window for fine-tuning in this model, but not a too small one.
	
\begin{figure}
		\centering
		\includegraphics[width=1.1\columnwidth]{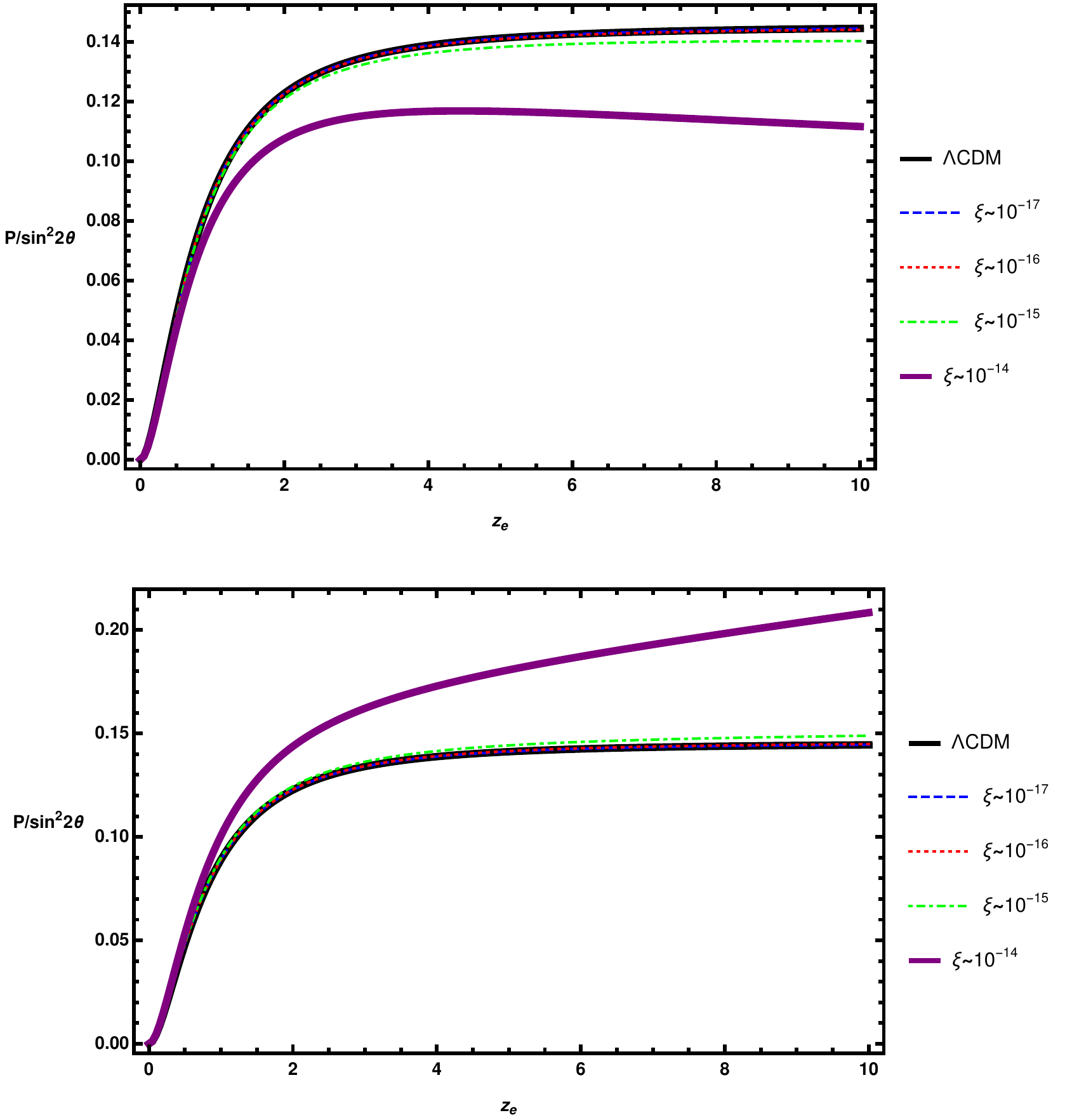}
		\caption{Evolution of the neutrino oscillation probability with redshift in the case of $\Lambda$CDM(solid black line) and quintessence, for neutrino-quintessence coupling $\sim\mathcal{O}(10^{-17})$(dashed blue), $10^{-16}$(dotted red), $10^{-15}$(dot-dash green) and $10^{-14}$(solid purple line). The parameters used are given in the text after eq.~\eqref{P_SA}, except for $\cos2\theta\sim0.2$ and $\Delta_m^2$, which is $2.51\times10^{-3}$eV${}^2$(upper panel) for normal hierarchy, and $-2.56\times10^{-3}$eV${}^{2}$(lower panel) for the inverted one.}
		\label{fig:Plot_PQ_NH_IH}
	\end{figure}

	On another note, we can also explore how the results might change for different neutrino parameters. Unlike for $\Lambda$CDM,  the neutrino-quintessence interaction is affected by the value of $\Delta_m^2$ and its hierarchy, which is evident from the denominator of eq.\eqref{P_Q}. To see this, we plot in the upper panel of figure~\ref{fig:Plot_PQ_NH_IH} the probabilities shown in figure~\ref{fig:Plot_ProbQ_Chis}, but for $\Delta_m^2=2.51\times10^{-3}$eV${}^2$ (normal hierarchy), while in the lower panel we use $\Delta_m^2=-2.56\times10^{-3}$eV${}^2$ (inverted hierarchy), with $\cos2\theta\sim0.2$ for both. These values of $\Delta_m^2$ correspond to the difference between neutrino mass states 3 and 2 of the standard neutrino oscillation treatment~\cite{Review_Particle_Physics}.
	
	There are a few things to be noted from these plots. First, if our neutrino mass states are 3 and 2 from~\cite{Review_Particle_Physics}, it is more difficult to distinguish $\Lambda$CDM from the quintessence model considered here for coupling constants smaller than $10^{-14}$. This difficulty can be evaded once we consider the full 3-flavor neutrino oscillations, which will be done in future works. Our purpose here is merely to show that different DE models affect neutrino oscillations differently. Second, even when we include all three neutrino flavors, there will be a noticeable difference between the two hierarchies for larger values of the coupling($\sim10^{-14}$), as is apparent from the two panels of figure~\ref{fig:Plot_PQ_NH_IH}. Therefore, such a neutrino-quintessence interaction could require some fine-tuning to match future observations, which puts it at equal, or less, footing with $\Lambda$CDM\footnote{Unless the value of the coupling constant is derived from a more fundamental theory, which establishes a fixed distinction between this model and $\Lambda$CDM.}.
	
	
	\subsection{Observational Strategy} Let us now comment on the relationship between our findings and observable quantities. Note that, due to the fact that we are considering a two-flavor neutrino system, direct comparison with neutrino observations would not be very beneficial. Nevertheless, our main results, presented in figures~\ref{fig:Plot_P_L_Dl} to~\ref{fig:Plot_PQ_NH_IH}, do affect neutrino observations, and we will be exploring this in more detail for three-flavor neutrinos in future work.
	
The main quantities observed at neutrino observatories, such as IceCube~\cite{IceCube}, are neutrino fluxes. For instance, the electron neutrino flux, $F_{\nu_e}$ can be expressed as~\cite{NuFlux1,NuFlux2}:\begin{equation}
		F_{\nu_e}=\sum_{\alpha=e,\mu\tau}^{}P_{\nu_{\alpha}\rightarrow\nu_e}F^0_{\nu_{\alpha}},
		\end{equation}where $F^0_{\nu_{\alpha}}$ is the flux of neutrinos with flavor $\alpha$ at the source. It is in this expression that our results could affect neutrino observations. The interaction of spinor neutrinos with curved spacetime will alter this expression through the transition probability $P_{\nu_{\alpha}\rightarrow\nu_e}$. More specifically to our case, depending on which DE model is considered, eqs.~\eqref{P_Lambda} and~\eqref{P_Q} will give different $P_{\nu_{\alpha}\rightarrow\nu_e}$ as a function of redshift, and thus the neutrino flux detected will be different. Therefore, by calculating the neutrino flux for each DE model, and compare it with observations, one can distinguish between these models.
	
	Another observational aspect worth mentioning is the experimental sensitivity available for such effects to be observed. We would like first to highlight that, when analyzing neutrino data, the usual emphasis is on the probability and flux's dependence on the neutrino's energy. However, in addition to this dependence, we are drawing attention here to the non-trivial effect of spacetime curvature on the observational results, which in an FLRW context translates into the dependence on the redshift. That is why in the analysis above a value for the energy of$\sim$10PeV has been chosen. Such a value is within reach of next generation neutrino detector IceCube-Gen2, which will have a 5 times better sensitivity than IceCube~\cite{IceCube-Gen2}.

	\section{Conclusions}
	\label{Sec:Conc.}
	In this paper, we have provided a proof of concept that distinct DE models can be distinguished using neutrino oscillations, particularly through the evolution of the oscillation probability with redshift. We first looked at a more general interaction between two-flavor neutrinos, as quantum spinor fields, and a classical scalar field in general spacetime. We focused on the interaction with a scalar field since it comprises a large class of models for DE, including $\Lambda$CDM, quintessence and some modified gravity scenarios (such as Horndeski theory~\cite{Horndeski:1974wa,Kobayashi:2019hrl}). Moreover, the interaction term considered includes a part that couples equally to both neutrino flavors, a flavor-global interaction, and another which is flavor dependent (see eq.~\eqref{Gen.Inter.}). The purpose is to examine the different effect these two terms have on the oscillation probability, which can be seen from the main result eqs. (\ref{Probability}, \ref{OmegaMinusOmega}) in section~\ref{Trans_Prob}.
	
	Furthermore, we applied this general formalism to two specific DE models, $\Lambda$CDM and quintessence, to produce observable contrast between them using neutrino oscillations. In the former model, we showed in figure~\ref{fig:Plot_P_L_Dl} the evolution of the oscillation probability with redshift when DE is a cosmological constant. We also show in that figure the oscillation probability in case of a direct substitution of the cosmological distance traveled by neutrinos(such as the luminosity distance) and their energy in the standard formula for neutrino oscillations eq~\eqref{Std_Prob}, what we called SA. The point of this contrast is that, if we detect $\nu_e$ from a type Ia supernova(SN), for example, and we want to calculate their flux (which depends on the $\nu_e$'s survival probability), SA would give a result $\sim50-80\%$(depending on the SN's redshift, see figure~\ref{fig:Plot_diff_L_Dl}) more than the actual value. This should be taken into account when doing neutrino observations in the future~\cite{IceCube-Gen2,Boser:2013oaa}.
	
	On the other hand, for quintessence, we looked at a derivative coupling between neutrinos and the scalar field that is motivated by symmetry breaking arguments~\cite{burgess:effectivelagrangian,Witten:2000dt,Mohapatra:1980ui}, which was referred to in~\cite{2018PDU....20...72S} as the DE${}_{\nu}$ model. This coupling, and others, have been already studied in~\cite{Khalifeh:2020bdg}, but we focused in this work on the observational consequences of such a coupling which, without it, $\Lambda$CDM and quintessence would be indistinguishable. In figure~\ref{fig:Plot_ProbQ_Chis}, we show the oscillation probability's evolution with redshift for the two models, with the DE${}_{\nu}$ coupling varied from $10^{-17}$ to $10^{-14}$eV${}^{-1}$. We also investigated the effect several $\Delta_m^2$ values from Particle Physics have on the probabilities, which in figure~\ref{fig:Plot_ProbQ_Chis} was produced assuming mass states 1 and 2 from~\cite{Review_Particle_Physics} as ours. This plot shows a clear distinction between $\Lambda$CDM and quintessence for several values of the DE${}_{\nu}$ coupling. However, if we consider states 2 and 3 as our mass states, it would become more difficult to distinguish the two DE models, unless the DE${}_{\nu}$ coupling is at least $\mathcal{O}(10^{-14})$, as seen in figure~\ref{fig:Plot_PQ_NH_IH}. Nevertheless, one can see from the two plots of figure~\ref{fig:Plot_PQ_NH_IH} a difference in the probability's evolution between the normal and inverted hierarchies, specially for high values of the DE${}_{\nu}$ coupling.
	
	In the future, we would like to generalize the present work further, by looking at the full three-flavor neutrino scenario, which should alleviate the distinction between mass states choice previously mentioned. However, we expect the difference between hierarchies' choice to remain even in this case, which prompts investigating its possible degeneracy with parameters of DE models. Furthermore, one could also look at another type of general interaction that could include other modified gravity models for DE, such as extended gravity~\cite{Capozziello:2011et} or higher dimensions~\cite{Maartens:2010ar}. Finally, with the advancement in neutrino detection techniques, we would expect these signals to appear in near future terrestrial experiments~\cite{IceCube-Gen2,Boser:2013oaa}, or perhaps underground lunar ones.
	
	\section{Acknowledgment}
	We would like to thank David Valcin and Nicola Bellomo for very helpful discussions. The work of ARK and RJ is supported by MINECO grant PGC2018-098866-B-I00 FEDER, UE. ARK and RJ acknowledge ``Center of Excellence Maria de Maeztu 2020-2023" award to the ICCUB (CEX2019- 000918-M).
	
	\bibliography{biblio}
	\bibliographystyle{elsarticle-num}
\end{document}